\documentclass[useAMS,usenatbib]{mn2e}

\usepackage{graphicx}
\usepackage{natbib}
\usepackage{amssymb, amsmath}

\newif\ifAMStwofonts
\AMStwofontstrue

  \newcommand{\bc}{\begin{center}}
  \newcommand{\ec}{\end{center}}
  \newcommand{\be}{\begin{equation}}
  \newcommand{\ee}{\end{equation}}
  
  \newcommand{\hMsun}{h^{-1}\>{\rm M_\odot}}
  \newcommand{\Mpc}{~h^{-1}~{\rm Mpc}}

  \newcommand{\vel}{\,{\rm km\,s^{-1}}}

\newcommand{\mincir}{\raise
  -2.truept\hbox{\rlap{\hbox{$\sim$}}\raise5.truept \hbox{$<$}\ }}
\newcommand{\magcir}{\raise
  -2.truept\hbox{\rlap{\hbox{$\sim$}}\raise5.truept \hbox{$>$}\ }}
\newcommand{\siml}{\raise
  -2.truept\hbox{\rlap{\hbox{$\sim$}}\raise5.truept \hbox{$<$}\ }}
\newcommand{\simg}{\raise
  -2.truept\hbox{\rlap{\hbox{$\sim$}}\raise5.truept \hbox{$>$}\ }}

\begin{document}

\title[The effects of baryons on the halo mass function]
  {The effects of baryons on the halo mass function}
\author[Weiguang Cui, et al.]
  {Weiguang Cui$^{1}$\thanks{wgcui@oats.inaf.it}, Stefano
    Borgani$^{1,2,3}$, Klaus Dolag$^{4,5}$, Giuseppe Murante$^6$,
    \\~\\
\LARGE{\rm Luca Tornatore$^{1}$}\vspace{0.2cm}\\ 
  $^1$ Astronomy Unit, Department of Physics, University of Trieste, via Tiepolo 11, I-34131 Trieste, Italy\\
  $^2$ INAF, Osservatorio Astronomico di Trieste, via Tiepolo 11, I-34131 Trieste, Italy\\
  $^3$ INFN -- National Institute for Nuclear Physics, Trieste, Italy\\
  $^4$ Universit\"atssternwarte M\"unchen, M\"unchen, Germany\\
  $^5$ Max-Planck-Institut f\"ur Astrophysik, Garching, Germany\\
  $^6$ INAF, Osservatorio Astronomico di Torino, Str. Osservatorio 25,
  I-10025, Pino Torinese, Torino, Italy}

\maketitle

\begin{abstract}
  We present an analysis of the effects of baryon physics on the halo
  mass function. The analysis is based on simulations of a
  cosmological volume having a comoving size of $410\Mpc$, which have
  been carried out with the Tree-PM/SPH GADGET-3 code, for a WMAP-7
  $\Lambda$CDM cosmological model. Besides a Dark Matter (DM) only
  simulation, we also carry out two hydrodynamical simulations: the
  first one includes non--radiative physics, with gas heated only by
  gravitational processes; the second one includes radiative cooling,
  star formation and kinetic feedback in the form of galactic ejecta
  triggered by supernova explosions. All simulations follow the
  evolution of two populations of $1024^3$ particles each, with mass
  ratio such to reproduce the assumed baryon density parameter, with
  the population of lighter particles assumed to be collisional in the
  hydrodynamical runs. We identified halos using a spherical
  overdensity algorithm and their masses are computed at three
  different overdensities (with respect to the critical one),
  $\Delta_c=200$, 500 and 1500.

  We find the fractional difference between halo masses in the
  hydrodynamical and in the DM simulations to be almost constant, at
  least for halos more massive than $\log (M_{\Delta_c} / \hMsun )\geq
  13.5$. In this range, mass increase in the hydrodynamical
  simulations is of about 4--5 per cent at $\Delta_c=500$ and $\sim 1$
  -- 2 per cent at $\Delta_c=200$. Quite interestingly, these
  differences are nearly the same for both radiative and
  non--radiative simulations.  Mass variations depends on halo mass
  and physics included for higher overdensity, $\Delta_c=1500$, and
  smaller masses. Such variations of halo masses induce corresponding
  variations of the halo mass function (HMF). At $z=0$, the HMFs for
  GH and CSF simulations are close to the DM one, with differences of
  $\mincir 3$ per cent at $\Delta_c = 200$, and $\simeq 7$ per cent at
  $\Delta_c=500$, with $\sim 10$ -- 20 per cent differences reached at
  $\Delta_c = 1500$. At this higher overdensity, the increase of the
  HMF for the radiative case is larger by about a factor 2 with
  respect to the non--radiative case. Assuming a constant mass shift
  to rescale the HMF from the hydrodynamic to the DM simulations,
  brings the HMF difference with respect to the DM case to be
  consistent with zero, with a scatter of $\mincir 3$ per cent at
  $\Delta_c=500$ and $\mincir 2$ per cent at $\Delta_c=200$.

  Our results have interesting implications  
   to assess uncertainties in the mass function calibration
  associated to the uncertain baryon physics, in view of cosmological
  applications of future large surveys of galaxy clusters.
\end{abstract}


\section{Introduction} \label{sec:1}

An accurate calibration of the halo mass function is at the hearth of
a range of cosmic structure formation studies, from the study of
galaxy formation through semi--analytical models
\citep[e.g.][]{Baugh06}, to the cosmological application of galaxy
clusters \citep[][]{allen_etal11}. Under the standard hierarchical
$\Lambda$CDM model, halos are formed from initial density peaks
through gravitational instability. The halo mass function (HMF
hereafter) is directly connected to the primordial density field.
Since the abundance of density peaks over a given mass scale $M$ only
depends on the r.m.s. value $\sigma_M$ of the linear fluctuation field
at that mass scale, the abundance of halos is expected to be universal
once expressed as a function of $\sigma_M$, as assumed by the
Press-Schechter approach based on the spherical collapse model
\citep{Press74} and in the ellipsoidal collapse extension by
\cite{Sheth99}.

Through the years, N--body simulations of large cosmological volumes
have been used to calibrate fitting functions for a universal HMF
\citep[e.g.][]{Jenkins01,Warren06,springel_nat05,Lukic07}. Thanks to the
progressive increase in the covered dynamic range of halo masses,
simulation results have been shown to predict subtle but sizable
deviations from universality of the mass function. For instance
\cite{Reed03} found that the universal mass function by \cite{Sheth99}
over-predicts the number of most massive halos found at $z>10$. This
result was confirmed by the subsequent analysis by \cite{Reed07}, who
pointed out that an even better fit for the mass function can be
obtained if it is allowed to depend not only on the linear r.m.s. 
overdensity, but also on the local slope of the linear power spectrum at the
relevant mass scale.
Using the spherical over-density (SO) algorithm to measure cluster
masses, \cite{Tinker08} combined different simulations to calibrate
the HMF, with masses measured at different overdensities. They found
significant deviations from non-universality, with a monotonic
decrease of halo abundance with increasing redshift, and provided
fitting functions to such deviation.  Besides confirming the
non-universal behaviour of the high end of the HMF,
\cite{crocce_etal10,Tinker08} also pointed out that using more accurate
second--order Lagrangian perturbation theory to set initial conditions
could is relevant for an accurate HMF calibration.
\cite{bhattacharya_etal11} analyzed the HMF for an extended suite of
simulations also including quintessence models with $w\ne -1$ for the
Dark Energy equation of state, and confirmed violation of universality
at the $\sim 10$ per cent level for the range of masses and redshift
covered by their simulations.

At least in principle, calibrating the mass function of DM halos with
great accuracy is just a technical problem to be tackled by extending
the dynamic range of simulations and the parameter space of considered
cosmological models. However, the back-reaction effects of baryons on
dark matter halos are known to impact on density profiles and,
therefore, on their mass. In turn, these back-reaction effects are
expected to depend on the detail of the physical processes, such as
radiative cooling, star formation and energy feedback from
astrophysical sources, which determine the distribution of baryons
within DM halos. \cite{Tinker08} included a non--radiative
hydrodynamical simulation of a large cosmological volume within the
large set of simulations that they analyzed, without however
discussing in detail the effect of baryons on the HMF. \cite{Rudd08}
compared the HMF computed for a DM-only simulation with those obtained
from the corresponding hydro-dynamical simulations, carried out with
an Adaptive Mesh Refinement (AMR) code both with non-radiative physics
and including the effect of gas cooling and star formation. After
computing masses at the viral radius, they found that the HMF for
non--radiative simulation is very close to the DM-only one, at least
in the mass range numerically resolved by both simulations. On the
other hand, the radiative simulation was found to produce a $\simeq
10$ per cent higher mass function, as a consequence of the higher
concentration halo concentration resulting from adiabatic contraction
\citep[e.g.][]{Gnedin04}. A significant increase of halo concentration
from adiabatic contraction is a well known consequence of
(over)efficient gas cooling
\citep[e.g.][]{Pedrosa09,Tissera10,Duffy10}. In line with this result
on halo concentration, also the total matter power spectrum in
radiative hydrodynamic simulations has been shown to have a higher
amplitude than for DM--only N--body simulations, small non--linear
scales $k > 1 h Mpc^{-1}$ \citep[][; Casarini et al., in
preparation]{Rudd08,Jing06,Daalen11,casarini_etal11}.  However, also
the simple case of non-radiative hydrodynamics has been suggested to
increase halo concentration, as a consequence of a redistribution of
energy between baryons and DM during halo collapse
\citep{Rasia04,Lin06}.  An increase of halo concentration turns into
an increase of halo masses, hence increasing the halo mass
function. \cite{zentner_etal08} suggested that the main effects of
baryons can be translated into a simple change of halo concentrations,
thereby resulting in a uniform relative shift of halo
masses. \cite{Stanek09} compared the cluster masses and mass functions
for a set of simulations including only DM, non radiative
hydrodynamics, as well as radiative runs with and without
pre--heating.  They reported for the pre--heated run an average
decrease of halo mass $M_{500}$\footnote{In the following, we will use
  the convention $R_{\Delta_c}$ to indicate the halo radius
  encompassing an average overdensity of $\Delta_c$ times the critical
  cosmic density $\rho_{\rm cr}(z)$. Accordingly, $M_{\Delta_c}$ is
  the halo mass contained within $R_{\Delta_c}$.} by $15$ percent with
respect to the non--radiative case, and $16$ percent halo mass
enhancement for simulation with cooling and star formation (CSF) with
respect to the DM simulation. These mass variations turn into
differences of the HMF of up to $\sim 30$ percent.
  \cite{Stanek09} based their analysis on two different sets of
  simulations, based on SPH and AMR codes, also using slightly
  different choices for the cosmological parameters. Furthermore,
  results for their CSF case were based only on re-simulations of the
  13 most massive halos identified in the original simulation volume.

In order to improve with respect to the current understanding of
baryon effects on the HMF, we present in this paper the analysis of
three cosmological simulations based on DM only, non--radiative
hydrodynamics and cooling, star formation and supernova (SN)
feedback. These simulations are carried out starting from the same
initial conditions and using the same Tree-PM/SPH code GADGET-3
\cite{springel05}. Resolution and box--size of our simulations are
adequate to cover the halo mass distribution over the range $\log
(M_{200} / \hMsun)\simeq (12.5-15)$ at $z=0$. Due to the inclusion of
hydrodynamics, the dynamic range covered by our simulations is in
general narrower than that accessible by N--body simulations used over
the last few years for precision calibrations of the HMF. For this
reason, the aim of this paper is not that of providing one more of
such calibrations, rather our goal is to asses in detail the impact of
baryons on the HMF.

This paper is organized as follows. In Section~\ref{sec:2}, we
describe the simulations. Section~\ref{sec:3} is devoted to the
presentation of the analysis method and results. After describing the
halo identification method based on spherical overdensity, we present
the results of our analysis in terms of mass variation of halos and
resulting effect on the HMF. Finally, we discuss our results and
present the main conclusions in Section~\ref{sec:4}.

\section{The simulations} \label{sec:2} We carry out simulations of a
flat $\Lambda$CDM cosmology with $\Omega_m = 0.24$ for the matter
density parameter, $\Omega_b = 0.0413$ for the baryon contribution,
$\sigma_8=0.8$ for the power spectrum normalization, $n_s = 0.96$ for
the primordial spectral index, and $h =0.73$ for the Hubble parameter
in units of 100$\vel {\rm Mpc}^{-1}$. Initial conditions have been
generated at $z=49$ using the Zeldovich Approximation for a
periodic cosmological box with comoving size $L=410\Mpc$.  Initial
density and velocity fields are sampled by displacing, at redshift
$z=41$, the positions of two sets of $1024^3$ particles each,
according to the Zeldovich approximation, from unperturbed positions
located onto two regular grids which are shifted by half grid size
with respect to each other. Masses of the particles belonging to the
two sets have ratio such that to reproduce the cosmic baryon fraction,
with $m_1\simeq 3.54 \times 10^{9} \hMsun$ and $m_2\simeq 7.36 \times
10^{8} \hMsun$. In the DM-only simulation both particle species are
treated as collisionless, while in the hydro-dynamical simulations
$m_2$ provides the mass of gas particles. We emphasize that this
prescription to set initial conditions for the DM simulation ensures
that it starts exactly from the same sampling of density and velocity
field as its hydro--dynamical counterpart. Convergence of the mass
  function against changing initial redshift and effect of using
  second-order Lagrangian Perturbation Theory (2LPT) have been
  discussed by \cite{Tinker08} and \cite {crocce_etal10}. Although
  small but sizeable effects have been detected in the high--end of
  the mass function, the general result is that the effect of 2LPT is
  rather small for initial redshift and resolution relevant for our
  simulations. Furthermore, since our analysis is focussed on the
  relative effect induced by the presence of baryons, we expect the
  main conclusions not to be affected by increasing the accuracy in
  the computation of displacements in the generation of initial
  conditions.

Simulations are carried out using the TreePM-SPH code {\small
  GADGET-3}, an improved version of the {\small GADGET-2} code
\citep{Springel05a}. In {\small GADGET-3} domain decomposition is
performed by allowing disjointed segments of the Peano--Hilbert curve
to be assigned to the same computing unit, thus turning into a
significant improvement of the work--load balance when run over a
large number of processors. Gravitational forces have been computed
using a Plummer--equivalent softening which is fixed to
$\epsilon_{Pl}=7.5h^{-1}$ physical kpc from $z=0$ to $z=2$, and fixed
in comoving units at higher redshift.

\begin{figure}
\includegraphics[width=8.5truecm]{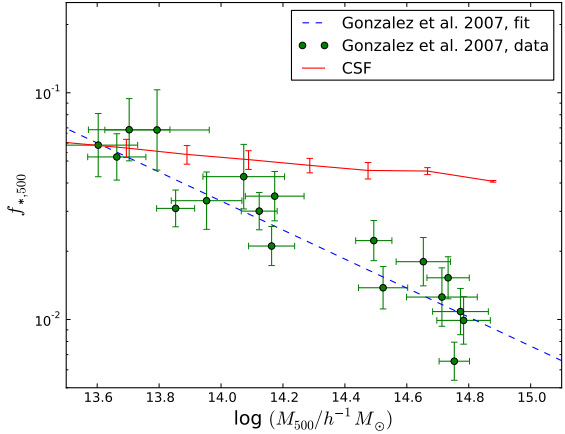}
\caption{Comparison between observational results and simulations on
  $f_*$, defined as the stellar mass fraction within $R_{500}$, for
  groups and clusters of galaxies. Circles with errorbars are the
  observational results for the clusters and groups analysed by
  \protect\cite{gonzalez_etal07}, with the dashed line showing the
  best-fit linear regression to these data points. The continous line
  show the results for the halos identified in the CSF
  simulations. Errorbars in this case refer to the r.m.s. scatter
  within each interval in $M_{500}$. }
\label{fig:fstar}
\end{figure}

Besides a DM-only simulation (DM hereafter), we also carried out two
hydrodynamical simulations. A non--radiative simulation only including
gravitational heating of the gas (GH hereafter) used 64 neighbours for
the computation of hydrodynamic forces, with the width of the B-spline
smoothing kernel allowed to reach a minimum value equal to half of the
gravitational softening. A second hydrodynamical simulation has been
carried out by including the effect of cooling and star formation (CSF
hereafter). In this simulation radiative cooling is computed for
non--vanishing metallicity according to \cite{sutherland_dopita93},
also including heating/cooling from a spatially uniform and evolving
UV background . Gas particles above a given threshold density are
treated as multi-phase, so as to provide a sub–resolution description
of the inter–stellar medium, according to the model described by
\cite{springel_hernquist03}. In each multi-phase gas particle, a cold
and a hot-phase coexist in pressure equilibrium, with the cold phase
providing the reservoir of star formation. Conversion of collisional
gas particles into collisionless star particles proceeds in a
stochastic way, with gas particles spawning a maximum of two
generations of star particles. The CSF simulation also includes a
description of metal production from chemical enrichment contributed
by SN-II, SN-Ia and AGB stars, as described by
\citep{tornatore_etal07}. Kinetic feedback is implemented by mimicking
galactic ejecta powered by SN explosions. In these runs, galactic
winds have a mass upload proportional to the local star-formation
rate. We use $v_w = 500\,{\rm km}\,s^{−1}$ for the wind velocity,
which corresponds to assuming about unity efficiency for the
conversion of energy released by SN-II into kinetic energy for a
Salpeter IMF. The feedback model included in the CSF simulation is
known not to be able to regulate overcooling, especially in large
cluster--sized halos \citep[e.g.][]{borgani_etal04}. To show
  this, we implement a consistent comparison in Figure \ref{fig:fstar}, 
  between observational results on
  the mass fraction in stars within $R_{500}$ (from
  \citealt{gonzalez_etal07}; see also
  \citealt{gonzalez_etal07,lagana_etal11}) and results obtained from
  the analysis of the clusters and groups identified in the CSF
  simulation. Quite apparently, simulations predict a decline of the
  stellar mass fraction as a function of cluster mass which is much
  milder that the observed one. As a result, massive systems in
  simulations are predicted to have an exceedingly high mass fraction
  in stars. Therefore, while none of the two hydrodynamical
simulations provides a fully correct description of the evolution of
baryons within DM halos, considering both the GH and the CSF runs one
should provide a useful indication of the impact of current
uncertainties in the description of baryon physics.

\section{Results} \label{sec:3}

\subsection{Halo identification}

The two most common methods for halo identifications simulations are
the one bases on the Friend-of-Friend (FoF) algorithm
\citep[e.g.][]{Davis85} and that based on the spherical overdensity
(SO) algorithm \citep{Lacey94}. The FoF halo finder has only one
parameter, $b$, which defines the linking length as $b l$ where
$l=n^{-1/3}$ is the mean inter-particle separation, with $n$ the mean
particle number density. In the SO algorithm there is also only one
free parameter, namely the mean density $\Delta_c~\rho_{crit}$
contained within the sphere within which halo mass is computed, with
$\rho_{crit}$ being the critical cosmic density. Each of the two halo
finders has its own advantages and shortcomings \citep[see more
details in][etc]{Jenkins01,White01,Tinker08}, and the difference of halo mass and
HMF defined by the two methods have been discussed in several
analysis, \citep[e.g.][]{White02,Reed03,Reed07,Cohn08,Tinker08,More11}.

In our analysis we apply the SO method, with masses measured at four
different overdensities corresponding to $\Delta_c=200$, 500 and
1500, thus ranging from the overdensity which characterize the whole
virialized region of halos up to the typical overdensity which is
accessible by Chandra and XMM--Newton X--ray observations.  Our halo
identification proceeds in two steps. In the first step we run a FoF
algorithm with linking length $b=0.16$ over the distribution of DM
particles (in the DM-only simulation the FoF is run over the
distribution of more massive particles). Then, we identify in each FoF
group the DM particle which corresponds to the minimum of the
potential. The position of this particle is taken to be the center of
the cluster from where to grow spheres whose radius is increased until
the mean density within it reaches the required overdensity
$\Delta_c$.  The mass $M_{\Delta_c}$ within this spherical region of
radius $R_{\Delta_c}$ is 
\be
M_{\Delta_c}=\frac{4}{3}{\pi}R^3_{\Delta_c} \Delta_c
{\rho}_{crit}(z)\,.  
\ee 

Since each halo is firstly identified starting from a FoF algorithm,
it inherits some FoF disadvantages. A well known potential problem
with FoF is that there are situations in which two halos are connected
through a bridge of particles. Since this halo is counted only once,
this could affect the number of SO halo number and the resulting mass
function. As discussed by \citep{Reed07}, this effect becomes more
important at high redshift and for poorly resolved low-mass
halos. Since we restrict our analysis to halos having a minimum mass
of $10^{13} \hMsun$, thus being resolved by at least $10^4$ particles)
and redshift $z\le 1$, and we use a linking length smaller than the
usually adopted value $b=0.2$, we expect the bias induced by using FoF
parent groups should be mitigated. 
 Since the FoF grouping is carried out using DM particles as
  primary particles, we expect halo bridging to affect in
  the same way the N-body and the different hydrodynamical
  simulations. Therefore, our main conclusions on the relative effect
  of baryons on the mass function should be left unchanged by the
  effect of using FoF groups as the starting point of the SO
  identification.
Finally, since the groups identified by FoF algorithm have by
definition no overlapping, we do not include in our identification of
SO halos any restriction to prevent such overlapping (see
\citealt{Tinker08} for a discussion on halo overlapping).

\begin{figure*}
\includegraphics[width=1.0\textwidth]{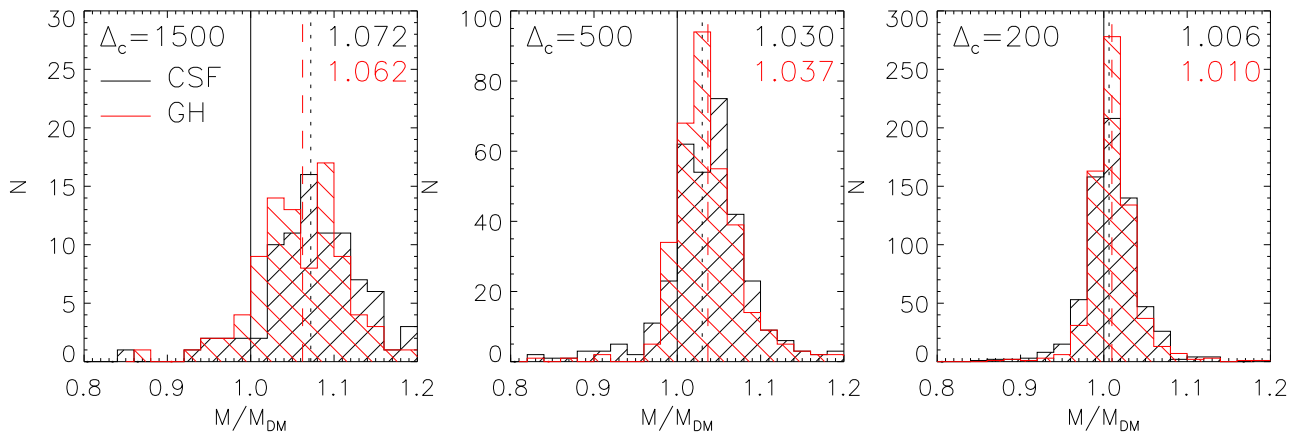}
\caption{The distribution of the mass ratios between the halos
  identified in each of the two hydrodynamical simulations to the 
  corresponding halos in the DM simulation. The
  three panels from left to right show the results at $\Delta_c = 1500, 500
  ,200$. Red and black continuous histograms show the results for
  $M_{GH}/M_{DM}$ and $M_{CSF}/M_{DM}$, respectively. In each panel,
  results are shown only for halos with $M_{\Delta_c} \geq 10^{14}
  \hMsun$. The solid vertical lines correspond to no mass variation,
  while the dashed and dotted vertical lines show the 
  mean values (which is shown on the right-top of each panel) of 
  $M_{GH}/M_{DM}$ and $M_{CSF}/M_{DM}$, respectively.}
\label{fig:mcomp}
\end{figure*}

\subsection{ Effect on halo mass and density profile}

We first focus on the impact that baryons have on the mass of
individual halos. To this purpose, we show in Figure \ref{fig:mcomp}
the distribution of the differences between halos identified in the
two hydrodynamical simulations and in the DM simulation, at different
overdensities. Results in this figure are shown for all the halos that
in the DM simulation have $M_\Delta\ge 10^{14}\hMsun$. To compare
halos masses, one has to identify a halo selected in the DM simulation
with its counterpart in each one of the hydrodynamical
simulations. The easiest way to perform this identification is to look
for the halos having the closest coordinates. While this procedure
provides a reliable identification of corresponding halos in two
different simulations for the most massive systems, it turns out not
to be accurate for poorer systems. In fact, besides affecting the mass
of halos, the presence of baryons also slightly alter the overall
dynamics and, therefore, the exact halo positions. In order to
overcome this difficulty we decided to follow a different procedure to
find in each of the hydrodynamical simulations the halos corresponding
to those identified in the DM run. For each halo in the DM simulation
we identify the Lagrangian region from where particles following
within its virial radius by $z=0$ come from. We then look in each of
the GH and CSF simulations for a halo that contains at least 60 per
cent of the particles coming from the same Lagrangian region. We
verified that the final results do not change significantly if we use
instead a more restrictive requirement to find instead 80 per cent of
the particles from the same Lagrangian region.

From Fig. \ref{fig:mcomp}, we see that significant mass differences,
of up to 20 per cent, are found for $\Delta_c=1500$, with the
distribution of such differences becoming narrower at
$\Delta_c=200$. Correspondingly, the mean value of the halo mass
increase induced by the presence of baryons decreases from $\simeq
6$--7 per cent at $\Delta_c=1500$ to $\simeq 3$--4 per cent at
$\Delta_c=500$, while being $\siml 1$ per cent at $\Delta_c=200$.
Furthermore, any differences between the two hydrodynamic runs is much
smaller than the difference that each of them has with respect to the
DM run. This result is in line with the weak sensitivity of the mass
function on the details of the baryon physics, as shown in
Fig. \ref{fig:HMF_D}. Even at the highest considered overdensity,
$\Delta_c=1500$, there is a small number of halos whose mass in the
hydrodynamical runs is smaller than in the DM run. The reason for this
is the different timing of merging of substructures in the different
simulations. This occasionally causes some of these substructures to
be found outside $R_{1500}$ while located within this radius in the DM
simulation.

\begin{figure*}
\includegraphics[width=1.0\textwidth]{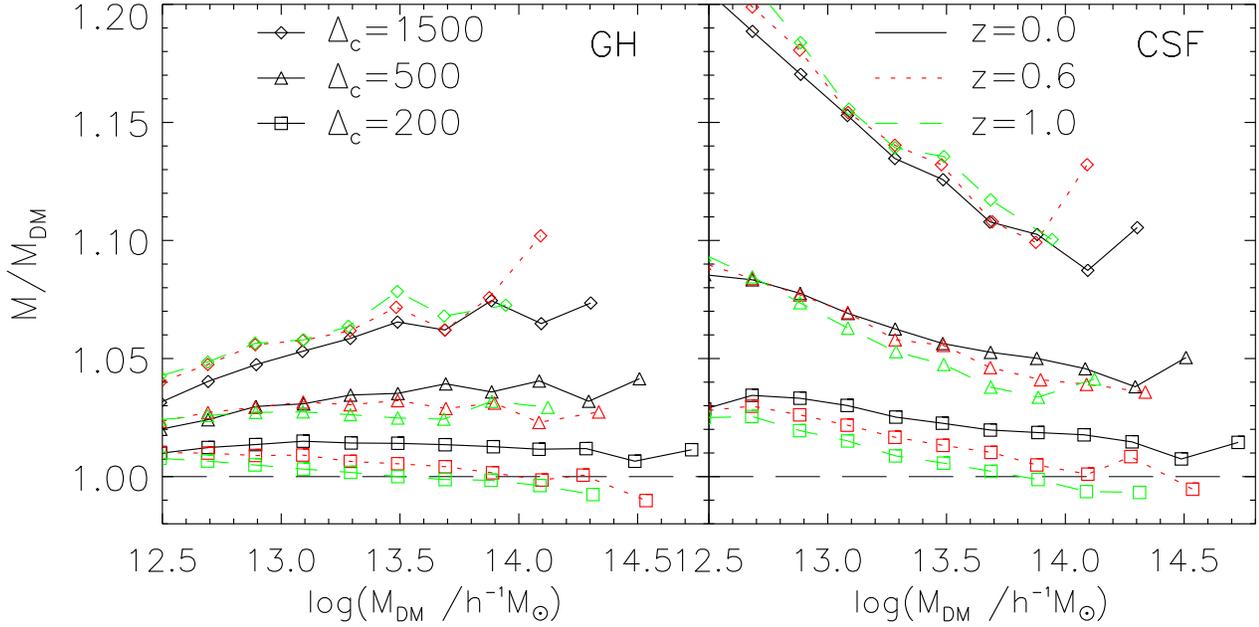}
\caption{The ratio between masses of halos identified in the
  hydrodynamical simulations and the corresponding halos from the DM
  simulation, as a function of halo mass in the DM run, $M_{DM}$. Left
  and right panels show the results for the GH and CSF run,
  respectively. In each panel different line styles, associated with
  different colors, corresponds to different redshifts.  Different
  symbols indicate instead different $\Delta_c$. For reference, the
  horizontal light long--dashed line correspond to no mass variation. }
\label{fig:dmf}
\end{figure*}

In order to quantify a possible mass dependence of this halo mass
difference, we show in Figure \ref{fig:dmf} the mean value of such a
difference for each mass bin where the mass function is computed.
Throughout our analysis, we use a fixed mass bin with width $\Delta
\log M = 0.2$. With such a narrow bin, the mass function suffers for
large sampling effect in the high--mass end, due to the exponential
dearth of the massive halo population. To overcome such sampling
effect, we merge mass bins containing less than 10 objects into the
adjacent lower mass bin. Each mass bin is then weighted proportionally
to the number of clusters it contains.

The increase of halo masses in both the GH and CSF simulations is to
good approximation independent of halo mass, at least for $\log
(M / \hMsun)\magcir 13.5$, at overdensities $\Delta_c=200$ and
500. Again, this shift in mass turns out to be similar in the
two hydrodynamical runs. It amounts to about 1--2 per cent at
$\Delta_c=200$ and $\simeq 4$ per cent at $\Delta_c=500$, in
line with the results shown in Fig. \ref{fig:mcomp}. The increasing
star formation efficiency in lower mass halos makes the mass increase
in the CSF simulation to be larger than for the GH case. This
difference between GH and CSF halo masses further increases for
$\Delta_c=1500$. At this overdensity we can not define a mass range
over which the increase of halo masses due to baryons is nearly
constant and independent of gas physics. 

\begin{figure*}
\includegraphics[width=1.0\textwidth]{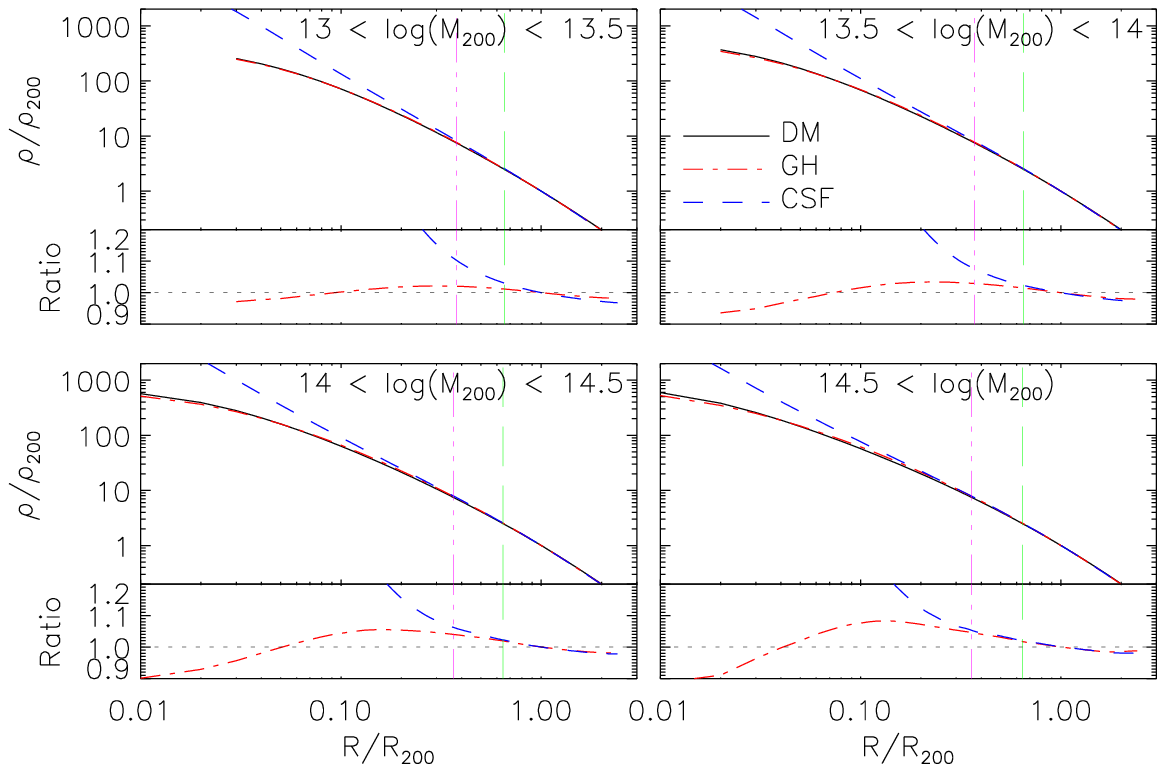}
\caption{Mean radial profiles of the density within a given
  radius. Density is expressed in units of $\rho_{200}$, which is the
  mean density within $R_{200}$. In each panel, solid (black),
  dot--dashed (red) and dashed (blue) curves correspond to the DM, GH
  and CSF runs, respectively. The two vertical lines mark the mean
  value of $R_{1500}$ and $R_{500}$ for the DM simulation (purple
  triple-dot--dashed and green long--dashed lines, respectively). The
  four panels correspond to four different mass ranges over which the
  mean profiles are computed, as indicated in the labels.}
\label{fig:den}
\end{figure*}

To better understand the origin of the mass difference between halos
identified in different simulations, we further show in Figure
\ref{fig:den} the radial profile of the mean total density for halos
identified in the three simulations. The four panels correspond to
different mass ranges. Since density is normalized to $\rho_{200}$,
i.e. the mean density within $R_{200}$, the profiles reach the unity
value for $R/R_{200}=1$. As for the halos identified in the GH
simulation (red dot-dashed curves), their profiles have small but
sizable differences with respect to the DM case (solid black
curves). At intermediate radii, $0.1\mincir R/R_{200}\mincir 1$ the GH
profiles lie above those of the DM simulation. This result, which
holds independent of the halo mass, is consistent with that found by
\cite{Rasia04} in their comparison of halo profiles from DM--only and
non--radiative hydrodynamical simulations. These authors argued that
the more concentrated density profiles in non--radiative simulations,
with respect to DM--only simulations, is the result of energy
redistribution between the DM and the baryonic component during halo
collapse \citep[see also][]{Lin06}. We postpone to a forthcoming paper
a detailed comparison between concentrations for halos identified in
DM and hydrodynamic simulations (Rasia et al. in preparation). It is
only at small radii, $R~\mincir 0.08R_{200}$, that gas pressure
support makes the total density profiles in the GH simulation slightly
flattening with respect to the DM simulation.

As for the radiative CSF simulation (blue dashed curves), the sinking
of cooled baryons, converted into stars, in the central halo regions
causes the already known effect of adiabatic contraction, with the
resulting steepening of the density profiles in these regions. A
comparison of the resulting profiles for the different mass ranges
indicates that this effect is more pronounced for halos of smaller
mass, consistent with the expectation that cooling is in fact more
efficient in lower mass halos, due to their higher concentration. The
effect of gas cooling is rather pronounced for $R\mincir
0.2R_{200}$. We note that the vertical purple line in Figure
\ref{fig:den} mark the mean value of $R_{1500}$ for halos of the DM
simulation. The corresponding value of $R_{1500}$ for the CSF
simulation is in fact slight larger than in the DM case. This
difference explains why halo masses in the CSF simulation are only
slightly larger than in the GH simulation already at $R_{1500}$.

\subsection{Effect on the Halo Mass Function}
\begin{figure}
\includegraphics[width=0.5\textwidth]{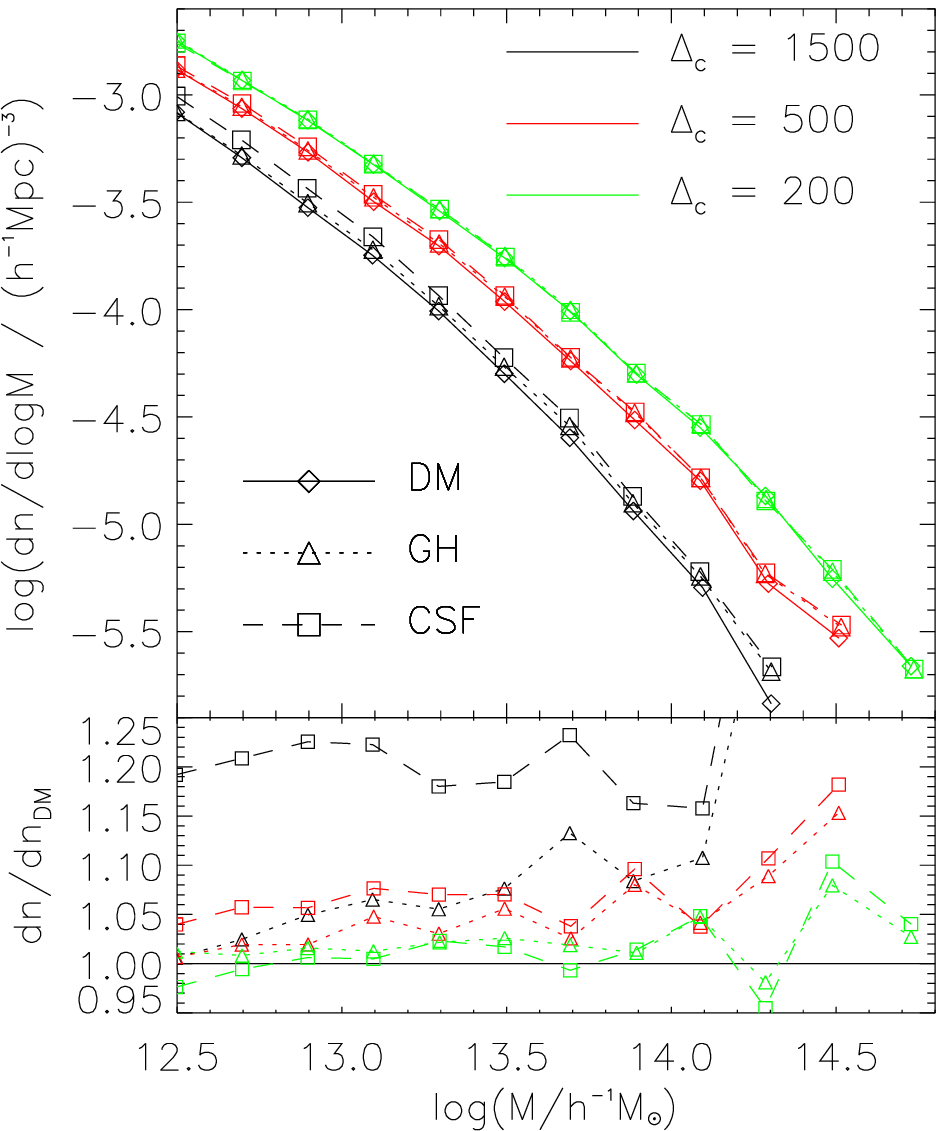}
\caption{The halo mass function for our simulations, with masses
  computed at different overdensities $\Delta_c$. Results for the DM
  simulations are shown with dotted curves, while results for the GH
  and CSF hydrodynamical simulations are shown with the dashed and
  dot--dashed curves, respectively.  Upper to lower curves correspond
  to result for $\Delta_c=200$, 500 and 1500 (green, red and black
  curves, respectively). The lower panel shows the ratio between the
  number of halos found in each mass bin for each of the two
  hydrodynamical simulation and the DM simulation. We apply a linear
  interpolation of the mass functions to compute the difference in the
  halo number exactly for the same mass values.}
\label{fig:HMF_D}
\end{figure}

In order to compute the mass function, we group SO halos within mass
bins having fixed width $\Delta \log M = 0.2$. Then, the mass
assigned to each bin is computed as the mean over all the halos
belonging to that mass bin. Whatever procedure one adopts to choose the
mass to be assigned to a given bin, it is clear that the binning
procedure introduce an uncertainty in the resulting mass function. As
discussed by \cite{Lukic07}, this uncertainty is negligible as log as
the bin width does not exceed $\Delta\log M=0.5$.

We show in Figure \ref{fig:HMF_D} the HMF for our three simulations,
computed for $\Delta_c = 200, 500, 1500$ (from upper to lower groups
of curves). To better emphasize the
mass variation induced by the presence of baryons, we show in the
lower panel the difference in the number of clusters within each mass
bin, between each of the two hydrodynamical simulations and DM
simulation. Due to the specific treatment of the last bin, its width
can be different for different simulations. Therefore, when we compare
the number of objects in such last bins, we rescale the cluster counts
within each of them by scaling it to the bin width in units of
$\Delta \log M = 0.2$.

In general, we find that the presence of baryons leads to an increase
of the HMF by an amount increasing as we move to more internal regions
at higher $\Delta_c$. In general, this variation is nearly independent
of mass, except possibly in the high mass end, beyond $\log(M /
\hMsun)\simeq 14.5$. This is the regime where exponential tail takes
place. Given the limited box size, the resulting limited statistics of
massive halos does not allow us to draw robust conclusions for such
high masses, especially when considering $\Delta_c=500$ and above. For
the GH non--radiative simulation the HMF increase is negligible at
$\Delta_c=200$, and amounts to $\mincir 3$ per cent at the largest
sampled masses. This difference increases to $\mincir 8$ per cent as
we move to $\Delta_c=500$, at least up to $\log (M / \hMsun)\simeq
14.5$. We note that at such overdensities the effect of introducing
baryons produce a variation of the HMF with respect to the DM
simulation which is larger than the difference between the GH and the
CSF run. This indicates that, while it is important to account for the
presence of baryons in the HMF calibration for $\Delta_c\mincir 500$,
the details of the physical processes regulating their evolution has a
minor impact. At a higher overdensity $\Delta_c = 1500$, the effect of
radiative physics is of increasing the HMF by about 20 per cent for
CSF run and around 10 per cent for GH run. This result is in line
with the expectation that a more concentrated density profile in the
presence of gas cooling \citep{Gnedin04,Pedrosa09,Tissera10,Duffy10}.

In general, our results for an increase of the HMF for the GH
simulation is in line with previous findings, also based on
non--radiative simulations, for an increase of halo concentration
induced by the presence of gas \citep{Rasia04,Jing06,Lin06,Rudd08}.
The effect of the physics of baryons becomes more important at
$\Delta_c=1500$. In their analysis of the cumulative mass function
\cite{Rudd08} found that the presence of non-radiative gas induces a
negligible HMF variation for masses estimated at the virial radius,
corresponding to $\Delta_C\simeq 100$ for their simulated
cosmology. While this result is in agreement with our, \cite{Rudd08}
find that the HMF increases by about 10 per cent when radiative
cooling and star formation are included. One possible reason for the
different effect of radiative physics in our analysis and in that of
\cite{Rudd08} could lie in the different efficiency of the feedback
included in the simulations. In our case we include a rather efficient
SN feedback, that could mitigate the effects of adiabatic
contraction. \cite{Stanek09} compared results for a non--radiative
simulation and for a pre--heated radiative simulations. They found
that at $\Delta = 500$ the latter predicts a HMF which is lower than
the former. The reason for this is that the fairly strong pre--heating
introduced in their simulation at $z=4$ devoid halos by a substantial
amount of gas, which is later prevented to re-accelerate in the
forming larger halos.

\begin{figure}
\includegraphics[width=0.5\textwidth]{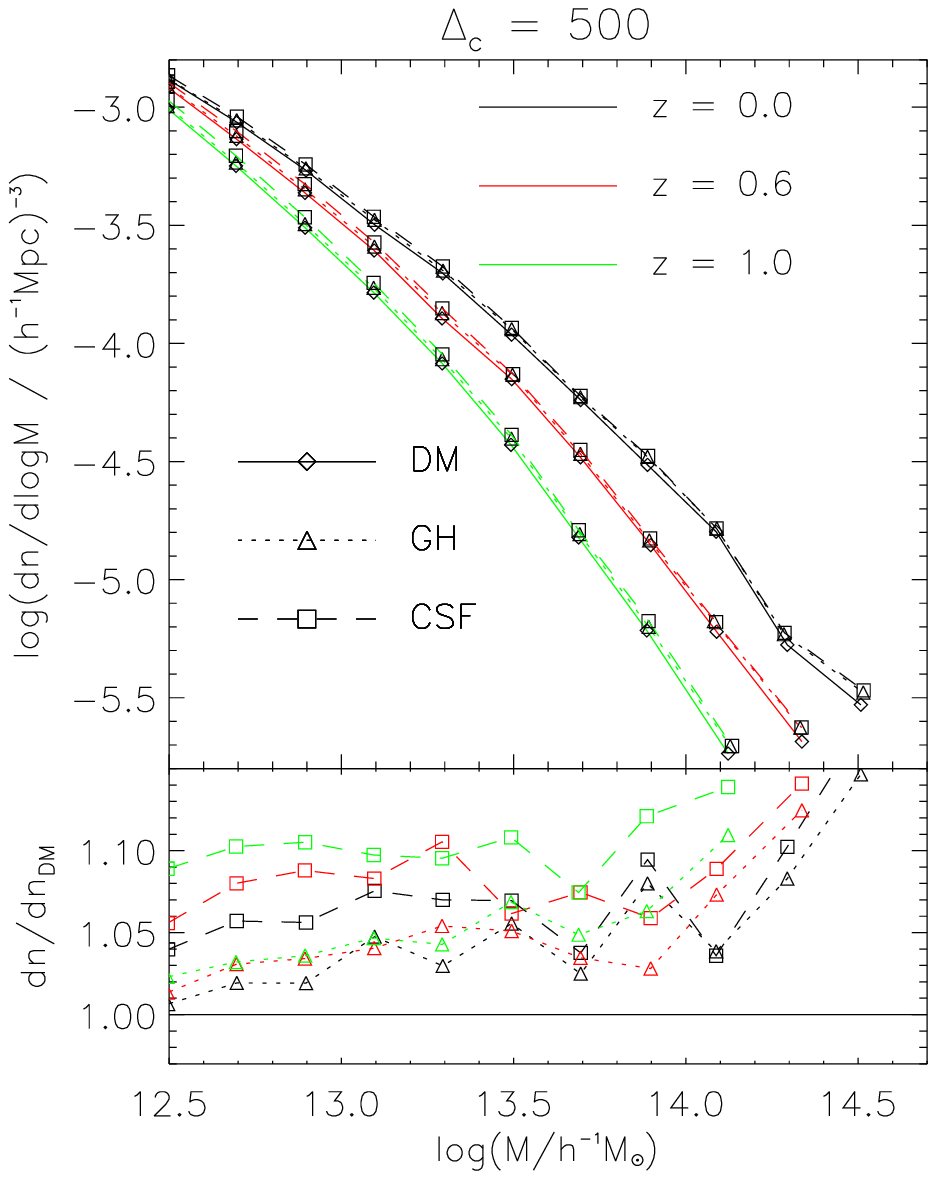}
\caption{The redshift evolution of the halo mass function for our
  simulations, with masses computed overdensity
  $\Delta_c=500$. Results for the DM simulations are shown with dotted
  curves, while results for the GH and CSF hydrodynamical simulations
  are shown with the dashed and dot--dashed curves, respectively.
  Upper to lower curves correspond to result for $z=0.00$, 0.58 and
  1.00 (black, red and green curves, respectively). The lower panel
  shows the ratio between the number of halos found in each mass bin
  for each of the two hydrodynamical simulation and the DM
  simulation.}
\label{fig:HMF_Z}
\end{figure}

\cite{Tinker08} investigated the redshift evolution of the mass
function computed at different values of $\Delta_c$ based on
simulations including only dark matter. They found that the abundance
of halos at a given $\log \sigma^{-1}$ monotonically decreases with
increasing z in the interval $[0, 2.5]$, where $\sigma$ is
r.m.s. variance of the linear density field smoothed on a given mass
scale. In the following we will discuss the impact that baryons in the
GH and CSF simulations have on the evolution of the HMF.

We show in Figure \ref{fig:HMF_Z} the evolution of the HMF at three
different redshifts, $z=0$, 0.6 and 1, for $\Delta_c =500$.  Similarly
to Fig. \ref{fig:HMF_D} we also show in the bottom panel the ratio
between the HMF from each of the two hydrodynamical simulations and
that of the DM simulation. In general, we find that the effect of
baryons on the MF is to good approximation independent of mass at all
redshifts for $\Delta_c = 500$. As for the non--radiative GH
simulation, the increase of the HMF with respect to the DM case is
always very small out to $z=1$, and $\mincir 8$ per cent. A more
significant effect is instead found for the radiative CSF
simulation. In this case the HMF increases by a larger amount at
progressively higher redshift, reaching the $\simeq 10$ per cent level
at $z=1$. This result agrees with the expectation that a more
efficient cooling takes place at higher redshift, which induces a
stronger effect on halo masses.

\begin{figure*}
\includegraphics[width=1.0\textwidth]{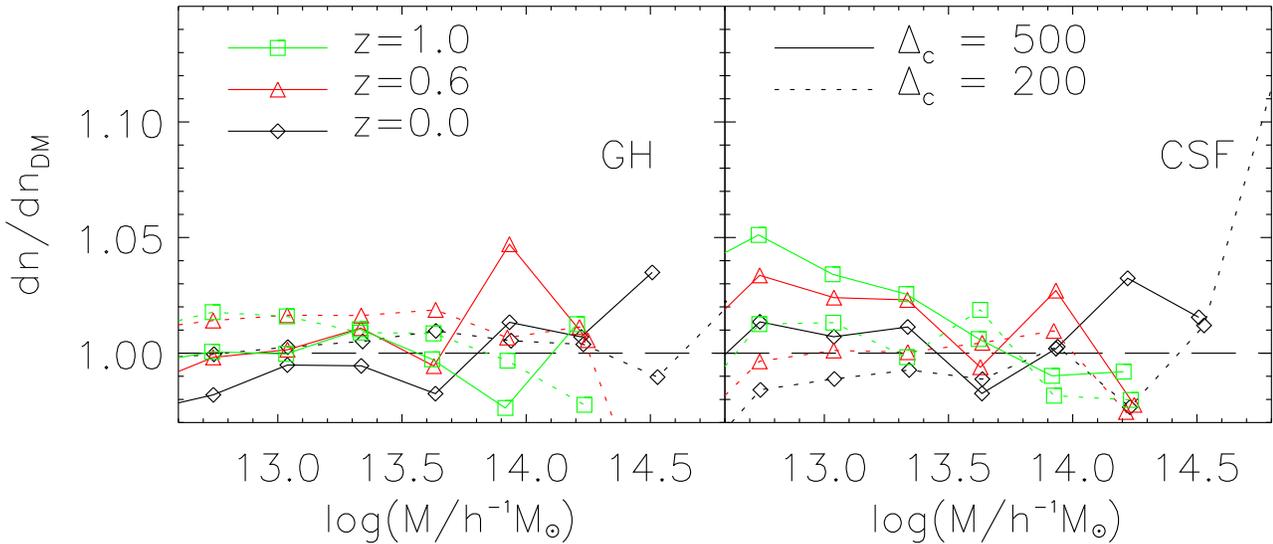}
\caption{The number difference after halo mass calibration. 
Different colorful lines show the redshift, while the two line styles,
solid and dotted represent $\Delta_c = 500, 200$, respectively. The left 
panel shows the corrected results for GH run, and the right is for CSF run.
}
\label{fig:correct}
\end{figure*}

\begin{table}
\begin{tabular}{l||cc||cc|}
\hline
Redshift & GH & & CSF &  \\
\hline
 & $\Delta_c=500$ & $\Delta_c=200$ & $\Delta_c=500$ & $\Delta_c=200$  \\
\hline
$z = 0.0$   & 1.037 & 1.012 & 1.049 & 1.016 \\
$z = 0.6$   & 1.028 & 1.000 & 1.044 & 1.005 \\
$z = 1.0$   & 1.028 & 0.997 & 1.040 & 0.999 \\
\hline
\end{tabular}
\caption{Mean values of the ratio between halo masses in the hydrodynamical
  and in DM simulations, at different redshifts and $\Delta_c$ values, for both the non--radiative (GH) and radiative (CSF) simulations. Such values
  have been computed by including only halos with$\log(M_{\Delta_c} / \hMsun)
  \geq 13.5$.}
\label{tab:meanmass}
\end{table}

Due to the complexity and partial knowledge of the baryonic process
taking place in galaxy formation, accurately calibrating the HMF from
hydrodynamical simulation may appear a challenging task. However,
owing to the results show in Fig. \ref{fig:dmf}, it turns out that
variation of halo masses in both GH and CSF simulations, with respect
to the pure DM case, is to good approximation independent of halo
mass, at least for $\log(M / \hMsun) \geq 13.5$ and $\Delta_c = 200$ and
500. Therefore, we can in principle attempt to account for the effect
of baryons in the HMF by a simple shift of halo masses, at least
within the mass range and for the $\Delta_c$ values for which is
approximation is expected to hold.  In Table \ref{tab:meanmass} we
report the average values of the mass ratio for halos identified in
the hydrodynamical and in the DM simulations, for different $\Delta_c$
and redshift values. This values, which are computed as an average
over all halos having mass $\log(M_{\Delta_c} / \hMsun) \geq 13.5$, are
then used to rescale the HMF from the hydrodynamical simulation.

In Figure \ref{fig:correct}, we show the ratio between the number of
halos of different mass found in the hydrodynamical and in the DM
simulations, after applying the mass shifts reported in Table
\ref{tab:meanmass}. A larger mass binning is used here, $\Delta \log M
= 0.3$, to reduce fluctuations associate to sampling noise. From the
left panel of Figure \ref{fig:correct}, the difference in the halo
number is now consistent with zero, with fluctuations around this
value of $\mincir 3$ percent for $\Delta_c = 500$. This result holds
independently of mass and redshift, at least for halos with $\log
(M_{\Delta_c} / \hMsun) \geq 13.5$. As expected, the correction is
less effective for smaller masses, owing to the larger mass difference
induced by baryonic effects in smaller halos.  Clearly, the correction
is less pronounced at $\Delta_c = 200$, owing to the smaller impact of
baryons at this overdensity. However, also in this case, correcting
the HMF according to a unique mass shift further reduces the
difference between hydrodynamical and DM simulations at the 1--2 per
cent level. The number difference for CSF run at $\Delta_c = 500$ is
also suppressed to unity for all halos with $\log (M / \hMsun) \geq
13.5$.

In conclusion, the results obtained from our analysis indicates that
the relative variation of halo masses due to baryon effects are always
within 5 per cent, for both non--radiative and radiative simulations,
also almost independent of redshift. This result holds for masses
computed at overdensity $\Delta_c=200$ and 500, and for halos having
mass at least comparable to that of a galaxy group. Correcting the
mass function with a constant mass shift in this mass range largely
accounts for the differences between hydrodynamical and DM
simulations.

\section{Discussion and conclusion} \label{sec:4}


In this paper we presented an analysis of the effect of baryons on the
calibration of the halo mass function (HMF). To this purpose, we
carried out one DM-only simulation (DM) and two hydrodynamical
simulations, a non--radiative one including only the effect of
gravitational gas heating (GH) and a radiative one including also the
effect of star formation and SN feedback in the form of galactic
ejecta. The three simulations, which are all based on the Tree-PM/SPH
GADGET-3 code \citep{springel05}, started from exactly the same
initial conditions and followed the evolution of $2\times 1024^3$
particles within a box having a comoving size of $410\Mpc$. Halos have
been identified using a spherical overdensity (SO) algorithm, and
results have been presented at three redshifts, $z=0$, 0.6 and 1. Halo
masses have been computed at different overdensities (with respect to
the critical one), $\Delta_c=200$, 500 and 1500. The main results of
our analysis can be summarized as follows.

\begin{description}
\item[1.] The fractional difference between halo masses in the
  hydrodynamical and in the DM simulations is found to be almost
  constant, at least for halos more massive than $\log (M_{\Delta_c} /
  \hMsun )\geq 13.5$. In this range, the mass increase in the
  hydrodynamical simulations is of about 4--5 per cent at
  $\Delta_c=500$ and 1--2 per cent at $\Delta_c=200$. Quite
  interestingly, these differences are nearly the same for the GH and
  the CSF simulations (see Fig.\ref{fig:dmf} and Table
  \ref{tab:meanmass}). Such relative mass variations can not be
  considered any more as constant at higher overdensity,
  $\Delta_c=1500$, and smaller masses. In these cases, mass difference
  markedly increases for smaller halos in the CSF simulation, while it
  decreases in the non--radiative GH simulation.
\item[2.] These variations of halo masses induce corresponding
  variations of the HMF (see Fig. \ref{fig:HMF_D}). At $z=0$, the HMFs
  for GH and CSF simulations are close to the DM one, with differences
  of $\mincir 3$ per cent at $\Delta_c = 200$, in line with the small
  correction in halo masses. Such a difference increases to $\simeq 7$
  per cent at $\Delta_c=500$ and reaches $\sim 10$--20 per cent at
  $\Delta_c = 1500$. At the latter overdensity, the increase in the
  HMF for the CSF run is larger by about a factor 2 with respect to
  the GH run. This result in line with the expectation that baryonic
  processes have a stronger impact in the central halo regions. At
  higher redshift, differences with respect to the DM HMF tend to
  increase, especially for $\Delta_c=1500$ (see \ref{fig:HMF_Z}) and
  for the CSF case. Again, this result agrees with the increase of
  cooling efficiency within halos at higher redshift.
\item[3.] Based on the above results, we showed that assuming a
  constant mass variation to rescale the HMF from the hydrodynamic
  simulations reduces the difference with respect to the DM case. We
  apply a uniform mass shift, calibrated for halo masses $\log (M \
  \hMsun) \geq 13.5$ for $\Delta_c = 200$ and 500. We verified that
  the difference between hydrodynamical and DM HMFs becomes
  negligible, with fluctuations around null of $\mincir 3$ per cent at
  $\Delta_c=500$. Even though mass variations are smaller at
  $\Delta_c=200$, we still find that a uniform mass rescaling gives a
  small but sizable reduction of the HMF difference also at this
  overdensity.
\end{description}


The future generation of large surveys of galaxy clusters, from
X--ray, optical and Sunyaev--Zeldovich observations, could provide
stringent constraints of cosmological parameters through the study of
the evolution of the mass function.  However, a necessary condition to
fully exploit the cosmological information content of such surveys is
that the theoretical mass function needs to be calibrated to a
precision better than 10 per cent \citep[e.g.][]{wu_etal10}. In this
respect, the results of our analysis have interesting implications to
gauge the uncertainty in the mass function calibration associated to
the uncertain baryon physics.

First of all, the HMF turns out to be less prone to such effects if
computed at $\Delta_c=500$, while they become more important and
likely difficult to model in detail at higher
overdensities. Furthermore, adopting a constant mass shift provides a
rather accurate correction to the HMF calibrated from DM simulations,
at least for halos having size or galaxy groups or larger. This result
holds for both the non--radiative (GH) and the radiative (CSF)
simulations, which have rather similar mass corrections at
$\Delta_c=500$. Since the CSF run only include SN feedback, but no AGN
feedback, clusters in this simulations still suffers for
overcooling.  Therefore, the GH and
CSF simulations should in principle bracket the case in which
the correct amount of baryons cools within DM halos. However, we
  note that \cite{Stanek09} found a slight decrease, rather than an
  increase of the HMF in simulations including an impulsive
  pre--heating. Since a phenomenological pre--heating only provides an
  approximate description of the astrophysical mechanisms regulating
  star formation, it would be interesting to repeat our analysis
also in the presence of a mechanism for AGN feedback that regulates
cooling in groups and clusters to the observed level
\citep[e.g.,][]{puchwein_etal08,fabjan_etal10,mccarthy_etal10}.

Another direction in which our analysis should be improved concerns
the size of simulation box, so as to better sample the population of
massive halos. Although our results indicate that a mass-independent
mass shift should be applied to account for baryonic effects, one may
wonder whether this prescription can be extrapolated to the most
massive halos, whose population is mostly sensitive to choice of the
cosmological model. Future development in supercomputing capabilities
will soon open the possibility to carry out hydrodynamical simulations
which will cover dynamic ranges comparable to those accessible by the
N--body simulations currently used to calibrate the halo mass
function.

\section*{Acknowledgements} 

The authors would like to thank anonymous referee for useful suggestions
and Pierluigi Monaco, Susana Planelles for valuable
discussions. Simulations have been carried out in CINECA (Bologna),
with CPU time allocated through ISCRA.  Weiguang Cui acknowledges a
fellowship from the European Commission's Framework Programme 7,
through the Marie Curie Initial Training Network CosmoComp
(PITN-GA-2009-238356). This work is partially supported by the
PRIN-INAF-2009 Grant ``Toward an Italian network for computational
cosmology'' and by the PD51-INFN grant.

\bibliography{paper,review_sb}

\begin{thebibliography}{}

\bibitem[\protect\citeauthoryear{{Allen}, {Evrard} \& {Mantz}}{{Allen}
  et~al.}{2011}]{allen_etal11}
{Allen} S.~W.,  {Evrard} A.~E.,    {Mantz} A.~B.,  2011, ArXiv e-prints

\bibitem[\protect\citeauthoryear{{Baugh}}{{Baugh}}{2006}]{Baugh06}
{Baugh} C.~M.,  2006, Reports on Progress in Physics, 69, 3101

\bibitem[\protect\citeauthoryear{{Bhattacharya}, {Heitmann}, {White},
  {Luki{\'c}}, {Wagner} \& {Habib}}{{Bhattacharya}
  et~al.}{2011}]{bhattacharya_etal11}
{Bhattacharya} S.,  {Heitmann} K.,  {White} M.,  {Luki{\'c}} Z.,  {Wagner} C.,
    {Habib} S.,  2011, \apj, 732, 122

\bibitem[\protect\citeauthoryear{{Borgani}, {Murante}, {Springel}, {Diaferio},
  {Dolag}, {Moscardini}, {Tormen}, {Tornatore} \& {Tozzi}}{{Borgani}
  et~al.}{2004}]{borgani_etal04}
{Borgani} S.,  {Murante} G.,  {Springel} V.,  {Diaferio} A.,  {Dolag} K.,
  {Moscardini} L.,  {Tormen} G.,  {Tornatore} L.,    {Tozzi} P.,  2004, \mnras,
  348, 1078

\bibitem[\protect\citeauthoryear{{Casarini}, {Macci{\`o}}, {Bonometto} \&
  {Stinson}}{{Casarini} et~al.}{2011}]{casarini_etal11}
{Casarini} L.,  {Macci{\`o}} A.~V.,  {Bonometto} S.~A.,    {Stinson} G.~S.,
  2011, \mnras, 412, 911

\bibitem[\protect\citeauthoryear{{Cohn} \& {White}}{{Cohn} \&
  {White}}{2008}]{Cohn08}
{Cohn} J.~D.,  {White} M.,  2008, \mnras, 385, 2025

\bibitem[\protect\citeauthoryear{{Crocce}, {Fosalba}, {Castander} \&
  {Gazta{\~n}aga}}{{Crocce} et~al.}{2010}]{crocce_etal10}
{Crocce} M.,  {Fosalba} P.,  {Castander} F.~J.,    {Gazta{\~n}aga} E.,  2010,
  \mnras, 403, 1353

\bibitem[\protect\citeauthoryear{{Davis}, {Efstathiou}, {Frenk} \&
  {White}}{{Davis} et~al.}{1985}]{Davis85}
{Davis} M.,  {Efstathiou} G.,  {Frenk} C.~S.,    {White} S.~D.~M.,  1985, \apj,
  292, 371

\bibitem[\protect\citeauthoryear{{Duffy}, {Schaye}, {Kay}, {Dalla Vecchia},
  {Battye} \& {Booth}}{{Duffy} et~al.}{2010}]{Duffy10}
{Duffy} A.~R.,  {Schaye} J.,  {Kay} S.~T.,  {Dalla Vecchia} C.,  {Battye}
  R.~A.,    {Booth} C.~M.,  2010, \mnras, 405, 2161

\bibitem[\protect\citeauthoryear{{Fabjan}, {Borgani}, {Tornatore}, {Saro},
  {Murante} \& {Dolag}}{{Fabjan} et~al.}{2010}]{fabjan_etal10}
{Fabjan} D.,  {Borgani} S.,  {Tornatore} L.,  {Saro} A.,  {Murante} G.,
  {Dolag} K.,  2010, \mnras, 401, 1670

\bibitem[\protect\citeauthoryear{{Gnedin}, {Kravtsov}, {Klypin} \&
  {Nagai}}{{Gnedin} et~al.}{2004}]{Gnedin04}
{Gnedin} O.~Y.,  {Kravtsov} A.~V.,  {Klypin} A.~A.,    {Nagai} D.,  2004, \apj,
  616, 16

\bibitem[\protect\citeauthoryear{{Gonzalez}, {Zaritsky} \&
  {Zabludoff}}{{Gonzalez} et~al.}{2007}]{gonzalez_etal07}
{Gonzalez} A.~H.,  {Zaritsky} D.,    {Zabludoff} A.~I.,  2007, \apj, 666, 147

\bibitem[\protect\citeauthoryear{{Jenkins}, {Frenk}, {White}, {Colberg},
  {Cole}, {Evrard}, {Couchman} \& {Yoshida}}{{Jenkins}
  et~al.}{2001}]{Jenkins01}
{Jenkins} A.,  {Frenk} C.~S.,  {White} S.~D.~M.,  {Colberg} J.~M.,  {Cole} S.,
  {Evrard} A.~E.,  {Couchman} H.~M.~P.,    {Yoshida} N.,  2001, \mnras, 321,
  372

\bibitem[\protect\citeauthoryear{{Jing}, {Zhang}, {Lin}, {Gao} \&
  {Springel}}{{Jing} et~al.}{2006}]{Jing06}
{Jing} Y.~P.,  {Zhang} P.,  {Lin} W.~P.,  {Gao} L.,    {Springel} V.,  2006,
  \apjl, 640, L119

\bibitem[\protect\citeauthoryear{{Lacey} \& {Cole}}{{Lacey} \&
  {Cole}}{1994}]{Lacey94}
{Lacey} C.,  {Cole} S.,  1994, \mnras, 271, 676

\bibitem[\protect\citeauthoryear{{Lagana}, {Zhang}, {Reiprich} \&
  {Schneider}}{{Lagana} et~al.}{2011}]{lagana_etal11}
{Lagana} T.~F.,  {Zhang} Y.-Y.,  {Reiprich} T.~H.,    {Schneider} P.,  2011,
  ArXiv e-prints

\bibitem[\protect\citeauthoryear{{Lin}, {Jing}, {Mao}, {Gao} \&
  {McCarthy}}{{Lin} et~al.}{2006}]{Lin06}
{Lin} W.~P.,  {Jing} Y.~P.,  {Mao} S.,  {Gao} L.,    {McCarthy} I.~G.,  2006,
  \apj, 651, 636

\bibitem[\protect\citeauthoryear{{Luki{\'c}}, {Heitmann}, {Habib}, {Bashinsky}
  \& {Ricker}}{{Luki{\'c}} et~al.}{2007}]{Lukic07}
{Luki{\'c}} Z.,  {Heitmann} K.,  {Habib} S.,  {Bashinsky} S.,    {Ricker}
  P.~M.,  2007, \apj, 671, 1160

\bibitem[\protect\citeauthoryear{{McCarthy}, {Schaye}, {Ponman}, {Bower},
  {Booth}, {Dalla Vecchia}, {Crain}, {Springel}, {Theuns} \&
  {Wiersma}}{{McCarthy} et~al.}{2010}]{mccarthy_etal10}
{McCarthy} I.~G.,  {Schaye} J.,  {Ponman} T.~J.,  {Bower} R.~G.,  {Booth}
  C.~M.,  {Dalla Vecchia} C.,  {Crain} R.~A.,  {Springel} V.,  {Theuns} T.,
  {Wiersma} R.~P.~C.,  2010, \mnras, 406, 822

\bibitem[\protect\citeauthoryear{{More}, {Kravtsov}, {Dalal} \&
  {Gottl{\"o}ber}}{{More} et~al.}{2011}]{More11}
{More} S.,  {Kravtsov} A.,  {Dalal} N.,    {Gottl{\"o}ber} S.,  2011, ArXiv
  e-prints

\bibitem[\protect\citeauthoryear{{Pedrosa}, {Tissera} \&
  {Scannapieco}}{{Pedrosa} et~al.}{2009}]{Pedrosa09}
{Pedrosa} S.,  {Tissera} P.~B.,    {Scannapieco} C.,  2009, \mnras, 395, L57

\bibitem[\protect\citeauthoryear{{Press} \& {Schechter}}{{Press} \&
  {Schechter}}{1974}]{Press74}
{Press} W.~H.,  {Schechter} P.,  1974, \apj, 187, 425

\bibitem[\protect\citeauthoryear{{Puchwein}, {Sijacki} \&
  {Springel}}{{Puchwein} et~al.}{2008}]{puchwein_etal08}
{Puchwein} E.,  {Sijacki} D.,    {Springel} V.,  2008, \apjl, 687, L53

\bibitem[\protect\citeauthoryear{{Rasia}, {Tormen} \& {Moscardini}}{{Rasia}
  et~al.}{2004}]{Rasia04}
{Rasia} E.,  {Tormen} G.,    {Moscardini} L.,  2004, \mnras, 351, 237

\bibitem[\protect\citeauthoryear{{Reed}, {Gardner}, {Quinn}, {Stadel},
  {Fardal}, {Lake} \& {Governato}}{{Reed} et~al.}{2003}]{Reed03}
{Reed} D.,  {Gardner} J.,  {Quinn} T.,  {Stadel} J.,  {Fardal} M.,  {Lake} G.,
    {Governato} F.,  2003, \mnras, 346, 565

\bibitem[\protect\citeauthoryear{{Reed}, {Bower}, {Frenk}, {Jenkins} \&
  {Theuns}}{{Reed} et~al.}{2007}]{Reed07}
{Reed} D.~S.,  {Bower} R.,  {Frenk} C.~S.,  {Jenkins} A.,    {Theuns} T.,
  2007, \mnras, 374, 2

\bibitem[\protect\citeauthoryear{{Rudd}, {Zentner} \& {Kravtsov}}{{Rudd}
  et~al.}{2008}]{Rudd08}
{Rudd} D.~H.,  {Zentner} A.~R.,    {Kravtsov} A.~V.,  2008, \apj, 672, 19

\bibitem[\protect\citeauthoryear{{Sheth} \& {Tormen}}{{Sheth} \&
  {Tormen}}{1999}]{Sheth99}
{Sheth} R.~K.,  {Tormen} G.,  1999, \mnras, 308, 119

\bibitem[\protect\citeauthoryear{{Springel}}{{Springel}}{2005a}]{springel05}
{Springel} V.,  2005a, \mnras, 364, 1105

\bibitem[\protect\citeauthoryear{{Springel}}{{Springel}}{2005b}]{Springel05a}
{Springel} V.,  2005b, \mnras, 364, 1105

\bibitem[\protect\citeauthoryear{{Springel} \& {Hernquist}}{{Springel} \&
  {Hernquist}}{2003}]{springel_hernquist03}
{Springel} V.,  {Hernquist} L.,  2003, \mnras, 339, 289

\bibitem[\protect\citeauthoryear{{Springel}, {White}, {Jenkins}, {Frenk},
  {Yoshida}, {Gao}, {Navarro}, {Thacker}, {Croton}, {Helly}, {Peacock}, {Cole},
  {Thomas}, {Couchman}, {Evrard}, {Colberg} \& {Pearce}}{{Springel}
  et~al.}{2005}]{springel_nat05}
{Springel} V.,  {White} S.~D.~M.,  {Jenkins} A.,  {Frenk} C.~S.,  {Yoshida} N.,
   {Gao} L.,  {Navarro} J.,  {Thacker} R.,  {Croton} D.,  {Helly} J.,
  {Peacock} J.~A.,  {Cole} S.,  {Thomas} P.,  {Couchman} H.,  {Evrard} A.,
  {Colberg} J.,    {Pearce} F.,  2005, \nat, 435, 629

\bibitem[\protect\citeauthoryear{{Stanek}, {Rudd} \& {Evrard}}{{Stanek}
  et~al.}{2009}]{Stanek09}
{Stanek} R.,  {Rudd} D.,    {Evrard} A.~E.,  2009, \mnras, 394, L11

\bibitem[\protect\citeauthoryear{{Sutherland} \& {Dopita}}{{Sutherland} \&
  {Dopita}}{1993}]{sutherland_dopita93}
{Sutherland} R.~S.,  {Dopita} M.~A.,  1993, \apjs, 88, 253

\bibitem[\protect\citeauthoryear{{Tinker}, {Kravtsov}, {Klypin}, {Abazajian},
  {Warren}, {Yepes}, {Gottl{\"o}ber} \& {Holz}}{{Tinker}
  et~al.}{2008}]{Tinker08}
{Tinker} J.,  {Kravtsov} A.~V.,  {Klypin} A.,  {Abazajian} K.,  {Warren} M.,
  {Yepes} G.,  {Gottl{\"o}ber} S.,    {Holz} D.~E.,  2008, \apj, 688, 709

\bibitem[\protect\citeauthoryear{{Tissera}, {White}, {Pedrosa} \&
  {Scannapieco}}{{Tissera} et~al.}{2010}]{Tissera10}
{Tissera} P.~B.,  {White} S.~D.~M.,  {Pedrosa} S.,    {Scannapieco} C.,  2010,
  \mnras, 406, 922

\bibitem[\protect\citeauthoryear{{Tornatore}, {Borgani}, {Dolag} \&
  {Matteucci}}{{Tornatore} et~al.}{2007}]{tornatore_etal07}
{Tornatore} L.,  {Borgani} S.,  {Dolag} K.,    {Matteucci} F.,  2007, \mnras,
  382, 1050

\bibitem[\protect\citeauthoryear{{van Daalen}, {Schaye}, {Booth} \& {Dalla
  Vecchia}}{{van Daalen} et~al.}{2011}]{Daalen11}
{van Daalen} M.~P.,  {Schaye} J.,  {Booth} C.~M.,    {Dalla Vecchia} C.,  2011,
  \mnras, pp 1079--+

\bibitem[\protect\citeauthoryear{{Warren}, {Abazajian}, {Holz} \&
  {Teodoro}}{{Warren} et~al.}{2006}]{Warren06}
{Warren} M.~S.,  {Abazajian} K.,  {Holz} D.~E.,    {Teodoro} L.,  2006, \apj,
  646, 881

\bibitem[\protect\citeauthoryear{{White}}{{White}}{2001}]{White01}
{White} M.,  2001, \aap, 367, 27

\bibitem[\protect\citeauthoryear{{White}}{{White}}{2002}]{White02}
{White} M.,  2002, \apjs, 143, 241

\bibitem[\protect\citeauthoryear{{Wu}, {Zentner} \& {Wechsler}}{{Wu}
  et~al.}{2010}]{wu_etal10}
{Wu} H.-Y.,  {Zentner} A.~R.,    {Wechsler} R.~H.,  2010, \apj, 713, 856

\bibitem[\protect\citeauthoryear{{Zentner}, {Rudd} \& {Hu}}{{Zentner}
  et~al.}{2008}]{zentner_etal08}
{Zentner} A.~R.,  {Rudd} D.~H.,    {Hu} W.,  2008, \prd, 77, 043507

\end{thebibliography}
\bibliographystyle{mn2e.bst}

\bsp

\label{lastpage}

\end{document}